\documentclass[twocolumn,nodate,showpacs,preprintnumbers,amsmath,amssymb,aps,prd]{revtex4}

\usepackage{graphicx}
\usepackage{amsmath,amssymb}

\def\be{\begin{equation}}
\def\ee{\end{equation}}
\def\ba{\begin{eqnarray}}
\def\ea{\end{eqnarray}}
\def\H{\mathcal{H}}

\def\h{\hat}
\def\t{\tilde}
\def\N{N}

\def\SU{{\rm SU}}

\def\g{\gamma}

\def\lp{{\ell}_{\rm Pl}}
\def\ub{\underline}

\def\e{{}^o\!e}
\def\w{{}^o\!\omega}

\begin{document}

\title{Quantum Nature of the Big Bang}
    \author{Abhay Ashtekar}
    \email{ashtekar@gravity.psu.edu}
    \author{Tomasz Pawlowski}
     \email{pawlowsk@gravity.psu.edu}
    \author{Parampreet Singh}
    \email{singh@gravity.psu.edu}
    \affiliation{Institute for Gravitational Physics and Geometry,
     Physics Department, The Pennsylvania State University, University Park, PA 16802,
    USA}

\begin{abstract}

Some long standing issues concerning the quantum nature of the big
bang are resolved in the context of homogeneous isotropic models
with a scalar field. Specifically, the known results on the
resolution of the big bang singularity in loop quantum cosmology
are significantly extended as follows: i) the scalar field is
shown to serve as an internal clock, thereby providing a detailed
realization of the `emergent time' idea; ii) the physical Hilbert
space, Dirac observables and semi-classical states are constructed
rigorously; iii) the Hamiltonian constraint is solved numerically
to show that the big bang is replaced by a big bounce. Thanks to
the non-perturbative, background independent methods, unlike in
other approaches the quantum evolution is deterministic across the
deep Planck regime.
\end{abstract}


\maketitle Thanks to the influx of observational data, recent
years have witnessed enormous advances in our understanding of the
early universe. To interpret the present data, it is sufficient to
work in a regime in which space-time can be taken to be a smooth
continuum as in general relativity, setting aside fundamental
questions involving the deep Planck regime. However, for a
complete conceptual understanding as well as interpretation of the
future, more refined data, these long-standing issues will have to
be faced squarely. Examples are: i) How close to the big-bang does
the smooth space-time of general relativity make sense? In
particular, can one show from first principles that this
approximation is valid at the onset of inflation? ii) Is the
big-bang singularity naturally resolved by quantum gravity? Or, is
some external input such as a new principle or a boundary
condition at the big bang essential? iii) Is the quantum evolution
across the `singularity' deterministic? In the Pre-big-bang and
Ekpyrotic scenarios, for example, the answer has been in the
negative \cite{other}. iv) If the singularity is resolved, what is
on the `other side'? Is there just a \emph{quantum foam} far
removed from any classical space-time, or, is there another large,
classical universe? The purpose of this letter is to summarize
results from recent analytical and numerical investigations within
loop quantum cosmology which address these and related issues.

Loop quantum gravity (LQG) is a background independent,
non-perturbative approach to quantum gravity \cite{alrev}. Loop
quantum cosmology (LQC) focuses on symmetry reduced models but
carries out quantization by mimicking the constructions used in
the full theory \cite{mbrev}. Results to date in this area fall in
two broad categories: a) Resolution of the big-bang singularity
using modifications of the {\it gravitational} Hamiltonian due to
quantum geometry \cite{mb1}; and, b) Phenomenological predictions
from effective equations that incorporate the modifications of the
\emph{matter} Hamiltonians due to quantum geometry (see, e.g.,
\cite{mb2,tsm}). As in the first category, we focus on the more
fundamental issues. While previous results showed that the LQC
evolution does not break down at the singularity, as pointed out
e.g. in \cite{bt}, they did not shed light on what happened
before. By constructing the missing conceptual and mathematical
infra-structure, we show that the universe has a classical
pre-big-bang branch, joined \emph{deterministically} to the
post-big-bang branch by the LQC evolution. Our detailed analysis
of the Planck regime also provides tools to test the validity of
assumptions underlying phenomenological predictions.

We will illustrate effects of quantum geometry on both the
gravitational and matter Hamiltonians through a simple example:
the spatially homogeneous, isotropic $k\!\!=\!\!0$ cosmologies
with a massless scalar field. Although the approach admits
generalizations, we focus on these models because a singularity is
unavoidable in their classical theory. The question is if it is
naturally tamed by quantum effects. The answer in the
`geometrodynamical' framework used in older cosmologies turns out
to be in the negative \cite{ck}. For example, if one begins with a
semi-classical state representing a classical universe at late
times and evolves it back via the Wheeler DeWitt equation, one
finds that it just follows the classical trajectory into the big
bang singularity\cite{aps}. In LQC, the situation is very
different. This may seem surprising at first. For, the system has
only a finite number of degrees of freedom and von Neumann's
theorem assures us that, under appropriate assumptions, the
resulting quantum mechanics is unique. However, for reasons we
will now explain, LQC does turn out to be qualitatively different
from the Wheeler-DeWitt theory \cite{abl}.

Because of spatial homogeneity and isotropy, one can fix a
fiducial (flat) triad $\e^a_i$ and its dual co-triad $\w_a^i$. The
$\SU(2)$ gravitational spin connection $A_a^i$ used in LQG has
only one component $c$ which furthermore depends only on time;
$A_a^i = c\,\, \w_a^i$. Similarly, the momentum $E^a_i$
canonically conjugate to $A_a^i$ ---representing a (density
weighted) triad--- has a single component $p$;\, $E^a_i = p\,(\det
\w)\, \e^a_i$. $p$ is related to the scale factor $a$ via $a^2 =
|p|$. However, $p$ is not restricted to be positive; under $p
\rightarrow -p$ the metric remains unchanged but the spatial triad
flips the orientation. The pair $(c,p)$ is canonically conjugate:
$\{c,\, p\} = (8\pi G \g/3)$, where $\g$ is the Barbero-Immirzi
parameter.

Quantization is carried out by closely mimicking the procedure
used in \emph{full} LQG \cite{abl}. There, the elementary
variables which have unambiguous operator analogs in quantum
theory are the \emph{holonomies} $h$ of connections $A_a^i$ and
the (smeared) triads $E^a_i$. Now, background independence leads
to a surprisingly strong result \cite{lost}: in essence, the basic
operator algebra generated by holonomies and triads admits a
\emph{unique} irreducible, diffeomorphism covariant
representation. In this representation, there are operators
$\h{h}$ representing holonomies and $\h{E}$ representing (smeared)
triads. But there are no operators representing connections
$A_a^i$ themselves. In the cosmological model now under
consideration, holonomies along a straight line of (oriented)
length $\mu$ (with respect to the fiducial triad $\e^a_i$) are
almost periodic functions of $c$ of the form $\N_{\mu}(c):= \exp
(i\mu c/2)$. (The $\N_{\mu}$ are the analogs of spin-network
functions in the full theory.) In the quantum theory, then, we are
led to a representation in which operators $\h{\N}_{\mu}$ and
$\h{p}$ are well-defined, but there is \emph{no} operator
corresponding to the connection $c$ itself (because the
1-parameter group $\h{\N}_{\mu}$ is not weakly continuous in
$\mu$). This \emph{new quantum mechanics is inequivalent to the
Wheeler-DeWitt theory already at a kinematical level}. In
particular, the gravitational part of the Hilbert space is now
$L^2(\bar{R}_{\rm Bohr}, d\mu_{\rm Bohr})$, the space of square
integrable functions on the Bohr compactification of the real
line, rather than the standard $L^2(R, d\mu)$ \cite{abl}. While in
the semi-classical regime LQC is well approximated by the
Wheeler-DeWitt theory, important differences manifest themselves
at the Planck scale. These are the hallmarks of quantum geometry
\cite{alrev,mbrev}.

The new representation also leads to a qualitative difference in
the structure of the Hamiltonian constraint operator: the
gravitational part of the constraint is a \emph{difference}
operator rather than a differential operator as in the
Wheeler-DeWitt theory. The derivation \cite{abl,aps} can be
summarized briefly as follows. In the classical theory, the
gravitational part of the constraint is given by $\int d^3x\,
\epsilon^{ijk} e^{-1} E^a_i E^b_j F_{ab\, k}$ where $e = |\det
E|^{1/2}$ and $F_{ab}^k$ is the curvature of the connection $A_a^i$.
The part of this operator involving triads can be quantized
\cite{abl} using a standard procedure introduced by Thiemann in
the full theory. However, since there is no operator corresponding
to the connection itself, one has to express $F_{ab}^k$ as a limit
of the holonomy around a loop divided by the area enclosed by the
loop, as the area shrinks to zero. Now, quantum geometry tells us
that the area operator has a minimum non-zero value, $\Delta$, and
in the quantum theory it is natural to shrink the loop only till
it attains this minimum. There are two ways to implement this idea
in detail. Here, we will use the one which has already appeared in
the literature \cite{abl,alrev,mbrev} although it has certain
drawbacks in the semi-classical regime, especially in more general
models. The second method will be discussed in the second of the
detailed papers \cite{aps}, which will also show that the quantum
bounce and deterministic evolution across the Planck regime
persist if the second and more satisfactory method is used. In
both cases, it is the existence of the `area gap' $\Delta$ that
leads one to a difference equation.

Let us represent states as functions $\Psi(\mu,\phi)$, where
$\phi$ is the scalar field and (modulo a fixed multiple of
$\lp^2$) the dimensionless real number $\mu$ is the eigenvalue of
$\h{p}$ \cite{abl}. Then in LQC the (self-adjoint) Hamiltonian
constraint is given by \cite{aps}
\ba {\partial^2_\phi\,\Psi}&=& [B(\mu)]^{-1}\,[C^+(\mu)
\Psi(\mu+4\mu_o,\, \phi) + C^o(\mu) \Psi(\mu, \, \phi)\nonumber\\
&+& C^-(\mu) \Psi(\mu-4\mu_o, \, \phi)] =: -\Theta \Psi(\mu,\phi)
\label{1}\ea
where $C^+(\mu)=(\pi G/9\mu_o^3)\,\mid |\mu+3\mu_o|^{3/2}\, -
|\mu+\mu_o|^{3/2}\mid $ ;\,\, $C^-(\mu) = C^+(\mu-4\mu_o)$;\,\,
$C^o(\mu) = -C^{+}(\mu) -C^-(\mu)$ and where $(6/8\pi
\gamma\lp^2)^{3/2}\,B(\mu)$ are the eigenvalues of the operator
$|\h{p}|^{-3/2}$ \cite{mbrev}. The fixed real number $\mu_o$ is
determined by the area gap; $(8\pi \gamma/6)\, \mu_o\, \lp^2 =
\Delta$.

Now, in each classical solution, $\phi$ is a globally monotonic
function of time and can therefore be taken as the dynamical
variable representing an \emph{internal} clock. In quantum theory,
even on-shell, there is no space-time metric. But since the
quantum constraint (\ref{1}) dictates how $\Psi(\mu,\phi)$
`evolves' as $\phi$ changes, it is convenient to regard the
argument $\phi$ in $\Psi(\mu,\phi)$ as `emergent time' and $\mu$
as the physical degree of freedom. A complete set of Dirac
observables is provided by the constant of motion $\h{p}_\phi$ and
operators $\h{\mu}|_{\phi_o}$ determining the value of $\mu$ at
the `instant' $\phi=\phi_o$.

Physical states are the (suitably regular) solutions to Eq
(\ref{1}). The map $\h\Pi$ defined by $\h\Pi\, \Psi(\mu, \phi) =
\Psi(-\mu, \phi)$ corresponds just to the flip of orientation of
the spatial triad (under which geometry remains unchanged);
$\h\Pi$ is thus a large gauge transformation on the space of
solutions to Eq. (\ref{1}). One is therefore led to divide
physical states into sectors, each providing an irreducible,
unitary representation of this symmetry. As one would expect,
physical considerations imply that we should consider the
symmetric sector, with eigenvalue +1 of $\h{\Pi}$ \cite{aps}.

To endow this space with the structure of a Hilbert space, we use
the `group averaging method' \cite{dm}. The technical
implementation of this procedure is greatly simplified by the fact
that the difference operator $\Theta$ on the right side of
(\ref{1}) is independent of $\phi$ and can be shown to be
self-adjoint and positive definite (on the Hilbert space
$L^2(\bar{R}_{\rm Bohr}, B(\mu) d\mu_{\rm Bohr})$. Since $\Theta$
is a difference operator, the resulting physical Hilbert space
$\H$ has sectors $\H_\epsilon$ which are superselected; $\H =
\oplus_\epsilon \H_\epsilon$ with $\epsilon \in [0,
2\mu_o]$. States $\Psi (\mu,\phi)$ in $\H_\epsilon$ (are symmetric
under the orientation inversion $\h{\Pi}$ and) have support on
points $\mu= \pm\epsilon + 4n\mu_o$. Let us consider a generic
$H_\epsilon$. (The small technical differences in the exceptional
cases are discussed in \cite{aps}; they do not affect the main
conclusions.) Wave functions $\Psi(\mu,\phi)$ solve (\ref{1}) and
are of positive frequency with respect to the `internal time'
$\phi$. Equivalently, they satisfy the `positive frequency' square
root of Eq (\ref{1}):
\be -i {\partial_\phi\,\Psi} = \sqrt{\Theta}\, \Psi \label{2}\ee
and the inner-product is given by:
\be \label{ip} \left<\!\Psi_1|\Psi_2\!\right>_{\rm phy} 
= \sum_{\mu \in
\{\pm\epsilon +4{\mu_o\mathbb{Z}}\}} B(\mu)\, \overline{\Psi_1}(\mu,
\phi) \Psi_2(\mu,\phi)\, , \ee
where as usual $\mathbb{Z}$ denotes the set of integers. On these states, the
Dirac observables act in the expected fashion:
\begin{subequations}\label{eq:obs}\begin{align}
\h{p}_\phi \Psi &= -i\hbar
{\partial_\phi\,\Psi}\\
 \h{\mu}|_{\phi_o}\,\, \Psi (\mu,\phi) &= e^{i
\sqrt{\Theta}(\phi-\phi_o)}\, \mu\, \Psi(\mu,\phi_o)
\end{align}\end{subequations}
One can also begin with the complete set of Dirac observables
(\ref{eq:obs}) and show that (\ref{ip}) is the unique inner product
which makes them self-adjoint.

To construct semi-classical states and for numerical simulations,
it is convenient to express physical states as linear combinations
of the eigenstates of $\h{p}_\phi$ and $\Theta$. We first note
that, for $\mu \gg \mu_o$, there is a precise sense \cite{aps,abl}
in which the difference operator $\Theta$ approaches the Wheeler
DeWitt differential operator $\ub{\Theta}$, given by
\be \big(\ub\Theta\,f\big) (\mu) = (16\pi G/3) \, \mu^{3/2}\,
(\sqrt{\mu} f')'\, .\ee
(Thus, if one ignores the quantum geometry effects, Eq (\ref{1})
reduces to the Wheeler-DeWitt equation $\partial^2_\phi\Psi =
-\ub{\Theta}\,\Psi$.) The eigenfunctions
\be \ub{e}_k(\mu)= (1/4 \pi)\,\, |\mu|^{1/4}\, e^{ik\ln |\mu|}\ee
of $\ub\Theta$ are labelled by a real number $k$ and its
eigenvalues are given by $\ub{\omega}^2 = (\pi G/3)(16 k^2+1)$.
The complete set of eigenfunctions $e_k(\mu)$ of the discrete
operator $\Theta$ is also labelled by a real number $k$ and
$e_k(\mu)$ are well-approximated by $\ub{e}_k (\mu)$ for $\mu \gg
\mu_o$ \cite{aps}. The eigenvalues $\omega^2 (k)$ of $\Theta$
increase monotonically with $|k|$. Finally, the $e_k(\mu)$ satisfy
the standard orthonormality relations $<\!e_k|e_k^\prime\!> =
\delta(k, k^\prime)$. A physical state $\Psi(\mu,\phi)$ can
therefore be expanded as:
\be \Psi(\mu,\phi) = \int_{-\infty}^{\infty}\! dk \t\Psi(k)\,\,
e_k^{(s)}(\mu)\,\, e^{i\omega(k)\phi}\,\ee
where $\t\Psi(k)$ is arbitrary (but suitably regular), $\omega(k)$
is positive and $e^{(s)}_k (\mu)= (1/\sqrt{2}) (e_k(\mu)+
e_k(-\mu))$. Thus, each physical state is characterized by a free
function $\t\Psi(k)$. (For proofs and subtleties see \cite{aps}.)

Since we have the explicit Hilbert space and a complete set of
Dirac observables, we can now construct states which are
semi-classical at late times ---e.g., now--- and evolve them
numerically `backward in time'. There are three natural
constructions to implement this idea in detail, reflecting the
freedom in the notion of semi-classical states. The main results
in all three cases are the same \cite{aps}. Here we will report on
the results obtained using the strategy that brings out the
contrast with the Wheeler DeWitt theory most sharply.

As noted before, $p_\phi$ is a constant of motion. For the
semi-classical analysis, we are led to choose a large value
$p_\phi^\star$ ($\gg \hbar$ in the classical $c\!=\!G\!=\!1$
units. In the closed model, for example, this condition is
necessary to ensure that the universe can expand out to a
macroscopic size.) Fix a point $(\mu^\star, \phi_o)$ on the
classical trajectory with $p_\phi = p_\phi^\star$ which starts out
at the big bang and then expands, choosing $\mu^\star \gg 1$. We
want to construct a state which is peaked at $(\mu^\star,
p_\phi^\star)$ at the initial `time' $\phi\!=\!\phi_o$ and follow
its `evolution' backward. Let $\t\Psi(k)$ be a Gaussian,  peaked
at a value $k^\star$ given by $p_\phi^\star= - (\sqrt{16\pi G
\hbar^2/3}) \, k^\star$. Set
\be\Psi(\mu,\phi_o) = \int_{\infty}^{\infty} \! dk\, \t\Psi(k)\,
\ub{e}_k(\mu)\, e^{i\ub{\omega}(k)(\phi_o-\phi^\star)}\ee
where $\phi^\star = - \sqrt{3/16\pi G}\, \ln |\mu^\star| + \phi_o$. It
is easy to verify that $\Psi(\mu,\phi_o)$ is the desired initial
data, sharply peaked at $p_\phi\! =\! p_\phi^\star$ and
$\mu\!=\!\mu^\star$. If evolved using the Wheeler-DeWitt analog
$-i\partial_\phi \Psi = \sqrt{\ub{\Theta}}\, \Psi$ of Eq
(\ref{2}), it would remain sharply peaked at the chosen classical
trajectory and simply follow it into the big-bang singularity
\cite{aps}. However, if it is evolved via (\ref{2}), the situation
becomes qualitatively different. The state remains sharply peaked
at the classical trajectory till the matter density reaches a
large critical value (which depends on $p_\phi^\star$), \emph{but
then bounces}, joining on to the `past' portion of a trajectory
which was classically headed towards the big crunch (see figures).
\begin{figure}
\begin{center}
\includegraphics[height=2in]{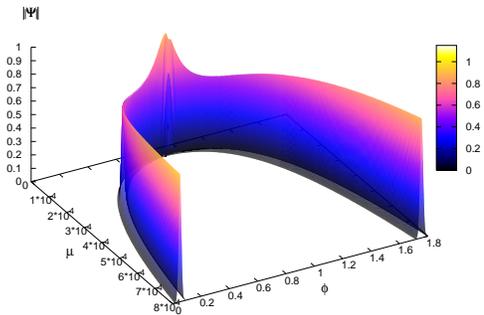}
\caption{The absolute value of the wave function $\Psi$ is plotted
as a function of $\phi$ and $\mu$ (whose values are shown in
multiples of $\mu_o$). For visualization clarity, only the values
of $|\Psi|$ greater than $10^{-4}$ are shown. Being a physical
state, $\Psi$ is symmetric under $\mu \rightarrow -\mu$.}
\label{Fig1}
\end{center}
\end{figure}
\begin{figure}
\begin{center}
\includegraphics[height=2in]{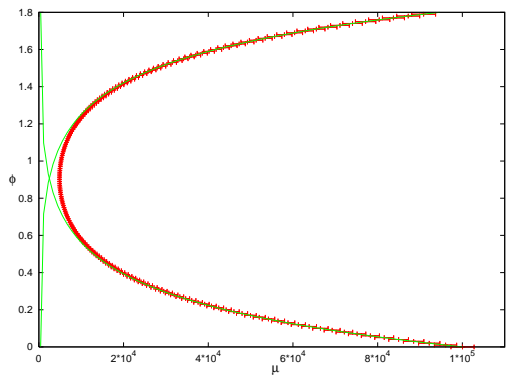}
\caption{The expectation values of Dirac observables
$\hat\mu|_{\phi}$ are plotted (in multiples of $\mu_o$), together
with their dispersions. They exhibit a quantum bounce which joins
the contracting and expanding classical trajectories marked by
fainter lines.} \label{Fig2}
\end{center}
\end{figure}

To ensure that these results are robust, a variety of numerical
simulations were performed to integrate Eq (\ref{1}) using the
adaptive step, 4th order Runge-Kutta method. Due to space
limitation, we will summarize only one of these. Here, we chose
$\epsilon\! =\! 2 \mu_o$ and initial data with $p_\phi^\star =
10^4 \sqrt{G\hbar^2}$,\, $\mu^\star= 10^5 \mu_o$ and the spread in
the Gaussian $\t\Psi(k)$ given by $\Delta p_\phi/p^\star_\phi =
7.5\times 10^{-3}$, where $\Delta{p_\phi}$ is the uncertainty in
$p_\phi$. The boundary of the numerical grid was chosen at
$1.5\mu^\star$ (where $|\Psi(\mu,\phi_o)| < 10^{-24}$.) On the
boundary, Eq (\ref{2}) was approximated by the Wheeler-DeWitt
equation and `outgoing wave' boundary conditions were imposed.
Results of the evolution exhibit a quantum bounce as shown in
figures 1 and 2. Away from the Planck regime the uncertainties in
the Dirac observables are essentially constant.

We conclude with a few remarks.\\
1. The main limitation of this analysis is the restriction to
homogeneity and isotropy. The approach can be readily extended to
incorporate anisotropic models and potentials for scalar fields.
However, incorporation of inhomogeneities has only just begun. The
hope is that the deterministic evolution of LQC will enable one to
evolve perturbations across the Planck regime. \\
2. The dramatic difference between the predictions of the Wheeler
DeWitt theory and LQC can be intuitively understood through
effective equations which can be derived from Eq (\ref{1}) using
certain approximations \cite{aps}. One finds that quantum geometry
(which is ignored in the Wheeler DeWitt theory) modifies the
Friedmann equations. The modifications are significant only in the
Planck regime and come with the sign
required to make gravity \emph{repulsive}. \\
3. A common feature with the early LQC papers is that we did not
have to introduce new physical input such as a boundary condition
at the singularity. We only asked that the quantum state be
semi-classical at late times. This is an observational fact rather
than a new theoretical input or a philosophical preference.
However, there are also notable differences from the existing LQC
literature. First, while much of the phenomenological work
\cite{mbrev} in LQC has incorporated quantum geometry effects only
on the matter Hamiltonian, here they were incorporated also in the
gravitational part. Second, we constructed the physical Hilbert
space, Dirac observables and semi-classical states, thereby
extracting physics of the Planck regime, going significantly
beyond the demonstration of singularity resolution. Specifically,
our results show that the quantum geometry in the Planck regime
serves as a `quantum bridge' between large classical universes,
one contracting and the other expanding. Finally, the idea that
the scalar field can be used as an internal clock has appeared
before, especially in \cite{kodama}. However, that analysis used
conventional quantum mechanics rather than the Bohr
compactification which descends from full LQG. Our final physical
Hilbert space is also different; it's construction is not
motivated by the Kodama state.

\textbf{Acknowledgment:} We would like to thank Martin Bojowald,
Jim Hartle, Wojciech Kaminski and Jerzy Lewandowski for
discussions. This work was supported in part by the NSF grants
PHY-0354932 and PHY-0456913, the Alexander von Humboldt
Foundation, and the Eberly research funds of Penn State.

\end{document}